\documentclass[]{article}
\pdfoutput=1 
\providecommand{\keywords}[1]{\textbf{\textit{Key words---}} #1}
\usepackage{lineno,hyperref}
\usepackage{amsmath}
\usepackage{amsfonts}
\usepackage{amssymb}
\usepackage{mathrsfs}
\usepackage[pdftex]{graphicx}
\usepackage{indentfirst}

\usepackage{lipsum} % Package to generate dummy text throughout this template

\usepackage[sc]{mathpazo} % Use the Palatino font
\usepackage[T1]{fontenc} % Use 8-bit encoding that has 256 glyphs
\usepackage{lmodern}
\usepackage{microtype} % Slightly tweak font spacing for aesthetics

\usepackage[hmarginratio=1:1,top=32mm,columnsep=30pt]{geometry} % Document margins
\usepackage[hang, small,labelfont=bf,up,textfont=it,up]{caption} % Custom captions under/above floats in tables or figures
\usepackage{booktabs} % Horizontal rules in tables
\usepackage{float} % Required for tables and figures in the multi-column environment - they need to be placed in specific locations with the [H] (e.g. \begin{table}[H])
\usepackage{hyperref} % For hyperlinks in the PDF

\usepackage{lettrine} % The lettrine is the first enlarged letter at the beginning of the text
\usepackage{paralist} % Used for the compactitem environment which makes bullet points with less space between them

\usepackage{abstract} % Allows abstract customization

\usepackage{titlesec} % Allows customization of titles
\renewcommand\thesection{\Roman{section}} % Roman numerals for the sections
\renewcommand\thesubsection{\Alph{subsection}} % Roman numerals for subsections
\titleformat{\section}[block]{\large\scshape\centering}{\thesection.}{1em}{} % Change the look of the section titles
\titleformat{\subsection}[block]{\large}{\thesubsection.}{1em}{} % Change the look of the section titles

\title{\vspace{-15mm}\fontsize{24pt}{10pt}\selectfont\textbf{Integral Control on Lie Groups}} % Article title

\author{
\large
{Zhifei Zhang$^{\dag\ddag}$\thanks{Corresponding author,~E-mail:~zhifei.zhang@ugent.be}~~~Alain Sarlette$^\ddag$~~~Zhihao Ling$^\dag$}\\[2mm] % Your name
\normalsize $^\dag$ \emph{Key Laboratory of Advanced Control and Optimization for Chemical Processes,}\\ \emph{Ministry of Education, East China University of Science and Technology,} \\\emph{No.130, Meilong Road, Shanghai 200237, China.} \\ % Your institution
\normalsize $^\ddag$ \emph{SYSTeMS research group, Ghent University,} \\\emph{Technologiepark Zwijnaarde 914, 9052 Zwijnaarde(Ghent), Belgium.} \\
\vspace{-5mm}
}

\date{}

\begin{document}

\maketitle

%\thispagestyle{fancy} % All pages have headers and footers

%----------------------------------------------------------------------------------------
%	ABSTRACT
%----------------------------------------------------------------------------------------

\begin{abstract}

\noindent  In this paper, we extend the popular integral control technique to systems evolving on Lie groups. More explicitly, we provide an alternative definition of ``integral action'' for proportional(-derivative)-controlled systems whose configuration evolves on a nonlinear space, where configuration errors cannot be simply added up to compute a definite integral. We then prove that the proposed integral control allows to cancel the drift induced by a constant bias in both first order (velocity) and second order (torque) control inputs for fully actuated systems evolving on abstract Lie groups. We illustrate the approach by 3-dimensional motion control applications.
\end{abstract}

%----------------------------------------------------------------------------------------
%	ARTICLE CONTENTS
%----------------------------------------------------------------------------------------

%\begin{multicols}{1} % Two-column layout throughout the main article text

\section{INTRODUCTION}

Exploiting the Lie group structure of rigid body motion to model robot configuration goes back to the Denavit-Hartenberg framework and its use for robotic arms \cite{Park1995lie}. Nowadays, the Lie group viewpoint has allowed to design common control methods for various mobile robot applications including satellite attitudes \cite{Sarlette2009autonomous, Hatanaka2012passivity, justh2005natural}, planar vehicles \cite{Sepulchre2008stabilization, justh2004equilibria},  submarines~\cite{LeonardPhD95,BulloPhD98}, surface vessels~\cite{Lefeber2003ships,Jiang2002ships}, quadrotor UAVs~\cite{Goodarzi2013geometric}, and their coordination~\cite{Sepulchre2008stabilization, Sarlette2009autonomous, Sarlette2010coordinated, Hatanaka2012passivity}. Lie groups involve a nonlinear configuration manifold where physical positions evolve, but with additional structure implying an almost linear viewpoint on the tangent bundle, where physical velocities evolve. The nonlinearity requires to adapt classical tracking and observer control tools. For example, a command proportional to configuration error must be defined as the gradient of an error function based on the distance-to-target along the manifold. The Lie group structure allows to systematically construct error functions from the relative configuration between system and target, e.g.~$\phi=\frac{1}{2}\text{tr}(I_{3\times 3}-Q_{\text{system}}^T Q_{\text{target}})$ for $Q_{\text{target}}, Q_{\text{system}}$ three-dimensional rotation matrices \cite{BulloPhD98,Bonnabel2008symmetry}. It also allows a canonical counterpart of Derivative control \cite{BulloPhD98}. However, in an attempt to generalize the Proportional-Integral (PI) and Proportional-Integral-Derivative (PID) controllers widely used for linear(ized) industrial control applications, the nonlinearity implies more fundamental issues for the integral control term.
%, and combined with other nonlinear control tools~\cite{Gao2002linear, Skjetne2004integral}. 
Indeed, simply integrating objects that belong to a manifold makes mathematically no sense. E.g.~a sum of rotation matrices gives in general an arbitrary square matrix of questionable use. Local linearization (retraction into a vector space) always allows a standard PI(D) control to be set up. This suggests that a proper extension might more globally recover the beneficial effect of integral control: rejecting with zero residual error a constant bias. The present paper proposes one way to extend PI(D) control to manifolds, and investigates more specifically how this rejects constant input biases on Lie groups.

In another recent approach, observers on Lie groups have been developed~\cite{Bonnabel2008symmetry, Lageman2010gradient} and applied to the estimation of bias in measurements~\cite{Khosravian2013bias}. The observer can also be used to estimate and compensate a bias in control commands. 
%We here have the same complementarity 
As in the linear case, the observer approach allows more accurate performance tuning, while the PID approach requires less model knowledge.

While this work was under review, the authors became aware of independent and concurrent work \cite{maithripala2014intrinsic} which the reader may want to consult for a complementary viewpoint.

%%%%%%%%%%%% NOTATION %%%%%%%%%%%%

\section{PRELIMINARIES AND NOTATION}\label{background}

\subsection{Dynamical systems on manifolds and Lie groups}\label{ssec:Liegroups}

Let $c(t)$ be the configuration at time $t$ of a system evolving on a nonlinear manifold $\mathcal{M}$ of finite dimension $d$. Its velocity $\dot{c} = \tfrac{dc}{dt}$ belongs to the tangent space to $\mathcal{M}$ at $c(t)$, which is a $d$-dimensional vector space $T_c\mathcal{M}$. The collection of such parameterized tangent spaces constitutes the tangent bundle $T\mathcal{M}$, a $2d$-dimensional manifold. The tangent space $T_{(c,\dot{c})}T\mathcal{M}$ to $T\mathcal{M}$ at $(c,\dot{c})$ is a vector space which, under canonical projection, contains the acceleration\footnote{Note that we are not speaking about Euler-Lagrange systems and possible curvature-induced accelerations here, we just define the spaces on which we work.} of the system on $\mathcal{M}$. 
A smooth vector field on $T\mathcal{M}$ (respectively on the acceleration-part of $TT\mathcal{M}$) defines a first-order (respectively second-order) system on $\mathcal{M}$ with well-defined integrated solution. In contrast, there is no intrinsic definition of what it would mean to mathematically integrate a position error which would be a function $c : \mathbb{R} \rightarrow \mathcal{M} : t \rightarrow c(t)$ over $t \in \mathbb{R}$.
For simplicity we identify tangent with cotangent space and let $\cdot$ be the scalar product between two vectors of $T_c\mathcal{M}$. The gradient $\text{grad}_c\phi \in T_c\mathcal{M}$ of $\phi : \mathcal{M} \rightarrow \mathbb{R}$ is defined such that $\text{grad}_c\phi \cdot v = \tfrac{d}{dt}\phi$ if $\tfrac{d}{dt}c = v$, for any $v \in T_c\mathcal{M}$. An element $v_1 \in T_{c_1}\mathcal{M}$ can be mapped to $T_{c_2}\mathcal{M}$ by a linear \emph{transport map}. The latter depends on a trajectory from $c_1$ to $c_2$, for which there are in general several canonical choices. The differential of a transport map on $T\mathcal{M}$ is an element of the acceleration class $TT\mathcal{M}$. The transport map is needed to compare tangent vectors (i.e.~velocities, accelerations) at different configurations.
%\vspace{3mm}
%If $v$ is taken by some transport map to another tangent space e.g.~at $c_1$, then we assume that $\text{grad}_c\phi$ is taken by the dual transport map to the corresponding cotangent space and denote it $\text{grad}_{c1}\phi$.

A Lie group $G$ is a smooth manifold with a group structure: a multiplication of $g,h\in G$ such that $g\cdot h\in G$, and an inverse $g^{-1}$ with respect to a particular $e\in G$ called identity, such that $g^{-1}\cdot g=g\cdot g^{-1}=e$. We denote the typical configuration on a group by $g$ instead of $c$. Lie groups feature canonical transport maps from $T_gG$ for any $g \in G$, to $T_eG \cong \mathfrak{g}$ the Lie algebra. The left-action transport map defines a left-invariant velocity $\xi^l = L_{g^{-1}}\frac{d}{dt} g$ and the right-action transport map a right-invariant velocity $\xi^r = R_{g^{-1}}\frac{d}{dt} g$. In practice, $\xi^l \in \mathfrak{g}$ and $\xi^r \in \mathfrak{g}$ often model the velocity expressed respectively in body frame and in inertial frame (although the correspondence is not always rigorous). Then left-invariant and right-invariant accelerations $\tfrac{d}{dt}\xi^l$ and $\tfrac{d}{dt}\xi^r$ can be defined on $\mathfrak{g}$ like for vector spaces.
The \emph{adjoint representation} $Ad_g$ is a linear $g$-dependent operator on the Lie algebra defined by $\xi^r = Ad_g \, \xi^l$ for any $dg/dt$. We have $Ad_{g^{-1}} = Ad_g^{-1}$, and $\tfrac{d}{dt}(Ad_g^{-1}) \chi^r = [\xi^l, Ad_g^{-1} \chi^r]$ for any constant $\chi^r \in \mathfrak{g}$ if $g$ moves according to $\xi^l = L_{g^{-1}}\frac{d}{dt} g$. Here we have introduced the Lie bracket, with property $[\xi_1,\xi_2] = - [\xi_2,\xi_1] \in \mathfrak{g}$ for all $\xi_1,\xi_2 \in \mathfrak{g}$. The gradient follows the dual mapping, e.g.~we note 
$\text{grad}^r\phi = Ad_{g^{-1}}^* \text{grad}^{l} \phi$ which indeed gives $\xi^r \cdot \text{grad}^r\phi = \xi^l \cdot \text{grad}^{l} \phi$. An important class of groups are compact groups with unitary adjoint representation, for which $Ad_g^* = Ad_{g^{-1}}$ or equivalently $[\xi_1,\xi_2] \cdot \xi_1 = 0$ for all $\xi_1,\xi_2 \in \mathfrak{g}$.
%\vspace{3mm}

\paragraph*{Example SO(3)} We represent the group of 3-dimensional rotations by $g$ a rotation matrix, group operations being the matrix counterparts, and $L_g$ the left matrix multiplication by $g$ of $\xi^l = [\omega^l]^{\wedge}$ a skew symmetric matrix in $\mathfrak{g} = \{S \in \mathbb{R}^{3\times 3} : S^T = -S \}$. The notation
$$\xi^l = [\omega^l]^\wedge = \begin{bmatrix}
    0 & -\omega_3 & \omega_2 \\
    \omega_3 & 0 & -\omega_1\\
    -\omega_2 & \omega_1 & 0\\
\end{bmatrix}
\Leftrightarrow [\xi^l]^{\vee} = \omega^l = \begin{bmatrix} \omega_1 \\ \omega_2 \\ \omega_3 \end{bmatrix}
$$
interprets $\omega^l$ as the angular velocity in body frame, $\omega^r = g\,\omega^l$ the angular velocity in inertial frame. For any matrix group, $\xi^r = g \xi^l g^{-1}$ and $[\xi^l_a,\xi^l_b] = \xi^l_a \xi^l_b - \xi^l_b \xi^l_a$. 

\paragraph*{Example SE(3)} The group of 3-dimensional rotations and translations is represented by $$g=\begin{bmatrix}
    R & p \\
    0_{1\times 3} & 1 \\
\end{bmatrix}$$
with $R \in SO(3)$ a rotation matrix and $p \in \mathbb{R}^3$ a translation vector. The group operations become matrix operations as for $SO(3)$, the elements of the Lie algebra write
$$\xi^l = g^{-1} \tfrac{d}{dt}g = \begin{bmatrix}
    [\omega^l]^{\wedge} & v^l \\
    0_{1\times 3} & 0 \\
\end{bmatrix}$$
with $v^l$ the translation velocity expressed in body frame. The group $SE(3)$ is not compact and hence its adjoint representation $Ad_g \xi^l = g \xi^l g^{-1}$ is not unitary: a large left-invariant velocity does not correspond to a large right-invariant velocity, and vice versa.

\subsection{Proportional and PD control on Lie groups}\label{ssec:refi}

PD controllers on manifolds and Lie groups have been previously proposed, see Introduction. Following a simplified version of \cite{BulloPhD98}, we define an error function $\phi(r^{-1} g)$ between current configuration $g(t)$ and target configuration $r(t)$. We make the typical assumption that $\phi(h)$ increases with the distance from $h$ to identity $e$, has a single local minimum $\phi(e)=0$ at the target, possibly (unavoidably on compact Lie groups) a set of other critical points.

For simplicity we assume $r$ to be fixed; feedforward can easily account for a moving $r(t)$, e.g.~by adding a term $\xi^l_{ff}=Ad_{g^{-1}r} \chi^l$ to the velocity command if $\tfrac{d}{dt}r = r\,\chi^l$.
In a first-order system,
$$\xi^l_p = -k_P \text{grad}^l\phi$$
is viewed as a \emph{proportional} feedback term. For a well-chosen $\phi$, the linearization of $\xi^l_p$ shall indeed be like proportional control for $r^{-1} g \simeq e$. In a second-order system
$$L_{g^{-1}}\tfrac{d}{dt}g = \xi^l \;\; , \quad \tfrac{d}{dt}\xi^l = F^l$$
with input torque/force $F^l$, the \emph{proportional} control is
$$F^l_p = -k_P \text{grad}^l\phi$$
and the \emph{derivative} control term is
$$F^l_d = -k_D \xi^l$$
(slightly more involved if $r$ was time-varying). A basic result of e.g.~\cite{BulloPhD98} (theorem 4.6) is that for fully actuated systems, both the first-order system with P-control and the second-order system with PD-control converge to the target, according to a Lyapunov function built around $\phi$.

In the following we show how to add integral control to this setting and recover this perfect convergence in presence of a constant input bias. In relation with this, we note that on Lie groups, a strong enough bias might not only prevent convergence close to the equilibrium, but even drive the system into a periodic motion. This is exemplified on the $N$-torus by weakly coupled Kuramoto oscillators with different natural frequencies~\cite{strogatz2000kuramoto}.

%%%%%%%%%%%% DEFINITION %%%%%%%%%%%%

\section{A DEFINITION OF INTEGRAL CONTROL in the PI / PID CONTEXT}\label{general methods}

In this section, we propose a general definition of integral control in the context of proportional or proportional-derivative control on nonlinear manifolds. In the next section, we specialize to Lie groups and prove how the proposed integral control allows to cancel the negative effect of constant biases.
We propose a simple \emph{intrinsic} way to define the integral control term on nonlinear manifolds, where the configuration error cannot be integrated:\vspace{2mm}

\noindent \textbf{Definition 1a:} The integral term $u_I$ for PI (respectively PID) control on a manifold is obtained as the integral of the P (respectively PD) control command $u_{PD}$.\vspace{2mm}

The spirit of this definition is to integrate the effort that the controller has been making so far. On a vector space, it is equivalent to the traditional definition as an integral of the output error. Indeed, we have:
\begin{itemize}
\item $u = k_P y + k_I \int y dt$ $\Leftrightarrow$ $u = k'_P y + k'_I \int (k'_P y) dt$ with $k'_P = k_P$, $k'_I = k_I/k_P$; and
\item $u = k_P y + k_D \dot{y} + k_I \int y dt$ $\Leftrightarrow$ $u = k'_P y + k'_D \dot{y} + k'_I \int (k'_P y + k'_D \dot{y}) dt$ with $k'_D = k_D$, $k'_P + k'_I k'_D = k_P$, $k'_I k'_P = k_I$, which has a solution as long as $k_p^2 \geq 4 k_I k_D$; incidentally, this is satisfied with equality for the Ziegler-Nichols tuning rules \cite{Astrom2010feedback}, yielding $k'_P = k_P/2$ and $k'_I = 2k_I/k_P = k_P/(2k_D)$.
\end{itemize}
Relaxing the spirit of strictly integrating the effort, one could also integrate the proportional and derivative terms with individual gains, i.e.~$u = k'_P y + k'_D \dot{y} + k'_{I,P} \int (k'_P y) dt + k'_{I,D} \int (k'_D \dot{y}) dt$. This would allow to cover the equivalent of linear controllers with any parameter values.% $k_p^2 < 4 k_I k_D$, if they do make sense for an application.

On manifolds, the control commands intrinsically belong to a vector space of dimension $d$, tangent to $\mathcal{M}$ or to $T\mathcal{M}$. As the system moves, the tangent space changes and in order to apply Definition 1a we must define how different vector spaces are related. %This is canonically done with a \emph{transport map}, hence:
\vspace{2mm}

\noindent \textbf{Definition 1b:} Explicitly, the integration of the control command for the integral term is given by:
$$ u_I(t) = \int_0^t \, T_{(x(\tau),x(t))}[u_{PD}(\tau)] \, d\tau \, ,$$
where $T_{(x(\tau),x(t))}$ is a transport map from the tangent space at the past configuration $x(\tau)$ to the tangent space at the current configuration $x(t)$. For a reasonably smooth choice of transport map, this can be written in differential form:
$$ \tfrac{d}{dt} u_I(t) = u_{PD}(t) + DT_{dx/dt}[u_I(t)]$$
where the linear map $DT_{dx/dt}[\cdot]$ accounts for the transport in the direction of the moving system.
\vspace{2mm}

The transport map from $x(\tau)$ to $x(t)$ may in general depend on the trajectory of the system. On Lie groups, there are two standard ways to define a transport map, related to left and right group actions (see Section \ref{ssec:Liegroups}). This allows for the following more detailed analysis.

%%%%%%%%%%%% CONVERGENCE %%%%%%%%%%%%

\section{CONVERGENCE ON LIE GROUPS}\label{general methods2}

In this section, in the spirit of a conceptual letter, we restrict our scope to PID control of pure integrators on Lie groups. We believe that like for linear systems there should be no major obstacle, in practical cases, to similarly prove stability in presence of more complicated dynamics (e.g.~nontrivial actuator transfer functions). We prove how the integral control stabilizes the system and at the same time completely rejects a constant input bias, illustrating the same prime effect as in linear systems. The stated conditions are only sufficient for convergence.

\subsection{Basic results}

A first-order system with our PI control and input bias $\xi^l_B$ in the left-invariant setting follows the dynamics:
\begin{eqnarray}
\label{eqmy1}
L_{g^{-1}}\frac{d}{dt} g &=& -k_p\, \text{grad}^l \phi + k_i\xi^l_i + \xi^l_B\,,\label{eq:1-st integral control}\\
\label{eqmy2}
\frac{d}{dt}\xi^l_i &=& -k_p\, \text{grad}^l \phi \;.\label{eq:1-st integral term}
\end{eqnarray}
We write $\xi^l$ instead of $u$, to emphasize that these are left-invariant velocities. Thanks to the left-invariant transport map, the integral control equation \eqref{eqmy2} takes a simple form. Following typical P and PD control strategies, we assume $\phi$ to have its only local minimum at the target $\phi=0$.\vspace{2mm}

\noindent \textbf{Proposition 1:} System \eqref{eq:1-st integral control},\eqref{eq:1-st integral term} converges globally to a set where $\text{grad}^l\phi=0$, according to a Lyapunov function 
\begin{equation}\label{eq:1st Lyapunov function}
V\,=\,\alpha\phi + \frac{1}{2}\beta\parallel k_i \xi_i^l + \xi^l_B\parallel^2\;,
\end{equation}
with $\alpha,\beta\,>\,0$. Only the equilibrium point with $\xi^l_i=-\xi^l_B/k_i$ and $\phi = 0$ is stable. The basin of attraction of that equilibrium can be increased to all $g$ for which $\phi(g) < \phi_c$ by taking $k_p k_i$ sufficiently large, if $\phi_c$ denotes the lowest value of $\phi>0$ for which $\phi$ has a critical point and we start at $\xi_i^l=0$ with a known bound on the bias $\Vert \xi^l_B \Vert^2 < B$.\vspace{2mm}

\noindent \emph{Proof.} Taking the time derivative of $V$ along the trajectory, reorganizing the terms and taking $\alpha\;=\;\beta k_pk_i$ gives
%\begin{equation}\nonumber
%\begin{split}
%\dot{V}\;=\;& \alpha \text{grad}^l\phi\cdot \xi^l + \beta(K_i\xi^l_i+\xi^l_B)\cdot K_i\dot{\xi^l_i} \\
%		 =\;&\alpha \text{grad}^l\phi\cdot(-K_p\text{grad}^l\phi+K_i\xi^l_i+\xi^l_B)\\
%		 &+\beta(K_i\xi^l_i+\xi^l_B)\cdot (-K_pK_i\text{grad}^l\phi)\\
%		 =\;&-\alpha K_p \Vert \text{grad}^l\phi \Vert^2\\
%		 &+(\alpha-\beta K_pK_i)\text{grad}^l\phi \cdot (K_i\xi^l_i+\xi^l_B)\;.
%\end{split}
%\end{equation}
$\dot{V}\;=\;-\alpha k_p (\text{grad}^l\phi)^2 \leq 0$. $V$ hence decreases everywhere except at the critical points of $\phi$. According to LaSalle's invariance principle, the system necessarily converges to an invariant set where $\dot{V}=0$, which must hence be contained in the set of critical points of $\phi$.

Only local minima of $V$ can be stable (since else a disturbance could push the system to a lower value of $V$, from which it would be unable to come back to its original state). This requires being at a local minimum of $\phi$, as the other term of $V$ only depends on other degrees of freedom, i.e.~velocities. By the assumption stated just before the Proposition, the minimum of $\phi$ reduces to the point where $\phi=0$. Staying at that point requires $\tfrac{d}{dt}g=0$ which characterizes the rest of the equilibrium.

Regarding the basin of attraction, $V(0) = V_0 < \alpha\, (\phi_c + \tfrac{\beta}{2\alpha} B)$ implies that the same bound holds for all $t>0$ as $V$ decreases over time. Any critical point except $\phi=0$ has $V \geq \alpha \, \phi \geq \alpha \phi_c$, while $V_0$ can be brought arbitrarily close to $\alpha \, \phi_c$ by increasing $\alpha/\beta$. Then for sufficiently large $\alpha/\beta = k_pk_i$, the system can at no future time reach any critical point of $\phi$ except $\phi=0$; and since we have proved above that the system converges to a critical point this concludes the proof.\hfill $\square$\\

In practice, Proposition 1 says that except for unstable trajectories, all the solutions should converge to the unique minimum $\phi=0$ corresponding to the target configuration. This is the same result as for P control without bias. The lack of a fully global result is due to the compactness of Lie (sub)groups which, unlike on vector spaces, precludes the existence of smooth $\phi$ with a unique critical point at $\phi=0$. Section \ref{applications} illustrates typical error functions $\phi$.\vspace{2mm}

A second-order system with our PID control and input bias $F^l_B$ in the left-invariant setting follows the dynamics:
\begin{eqnarray}
L_{g^{-1}}\frac{d}{dt}g & = & \xi^l \label{eq:2-nd integral control velocity}\\
\frac{d}{dt}\xi^l & = & -k_p \text{grad}^l \phi - k_d\xi^l + k_iF^l_i + F^l_B \label{eq:2-nd integral control force}\\
\frac{d}{dt}F^l_i & = & -k_p \text{grad}^l \phi - k_d\xi^l \;.\label{eq:2-nd integral term}
\end{eqnarray}
Both the bias and control terms now involve torques/forces, we emphasize this by writing $F$ instead of $u$.
\vspace{2mm}

\noindent \textbf{Proposition 2:} System \eqref{eq:2-nd integral control velocity}-\eqref{eq:2-nd integral term}, under the condition $K_i\,<\,K_d$, converges globally to an equilibrium set where $\text{grad}^l\phi=0$, $\xi^l=0$, $F^l_i=-F^l_B/K_i$, according to a Lyapunov function  
\begin{equation}\label{eq:2-nd Lyapunov function}
V\,=\,\alpha\phi + \frac{1}{2}\beta\parallel\xi^l\parallel^2 + \frac{1}{2}\gamma\parallel k_i (F^l_i-\xi^l) + F^l_B\parallel^2\;.
\end{equation}
with $\alpha,\beta,\gamma\,>\,0$. The stability, and basin of attraction for large $k_p$, hold as for Proposition 1.\vspace{2mm}% Within this equilibrium set, only the point with $\phi = 0$ is stable. Moreover, the basin of attraction of that equilibrium can be increased to all $g$ for which $\phi(g) < \phi_c$ by taking $K_p$ sufficiently large, if $\phi_c$ denotes the lowest value of $\phi>0$ for which $\phi$ has a critical point and we start at $\xi_i^l=0$ with a known bound on the bias $\Vert \xi^l_B \Vert^2 < B$.

\noindent \emph{Proof.} Inserting \eqref{eq:2-nd integral control velocity}-\eqref{eq:2-nd integral term} into the derivative of the proposed $V$ and taking $\alpha = \beta k_p$, we can get:
\begin{equation}\nonumber
\begin{split}
\dot{V} = & -\left(-\frac{1}{4\gamma k_i}\beta^2 + \left(k_d - \frac{1}{2}k_i\right)\beta-\frac{1}{4}\gamma k_i^3\right)\Vert\xi^l\Vert^2 \\
		 &-\left\Vert\frac{1}{2}\frac{\beta+\gamma k_i^2}{\sqrt{\gamma k_i}}\xi^l - \sqrt{\gamma k_i}(k_iF^l_i + F^l_B)\right\Vert^2\;.
\end{split}
\end{equation}
%\begin{equation}\nonumber
%\begin{split}
%\dot{V}\;&=\; \alpha \text{grad}^l\phi\cdot\xi^l \\
%&+ \beta\xi^l\cdot\dot{\xi^l} + \gamma(K_i(F^l_i-\xi^l) + F^l_B)\cdot(K_i(\dot{F^l_i}-\dot{\xi^l})) \\
%		 &=\;\alpha \text{grad}^l\phi\cdot\xi^l\\
%		 &+ \beta\xi^l\cdot(-K_p \text{grad}^l \phi - K_d\xi^l + K_iF^l_i + F^l_B)\\ 
%			&- \gamma K_i (K_i(F^l_i-\xi^l) + F^l_B)\cdot (-K_p \text{grad}^l\phi-K_d\xi^l + K_iF^l_i \\
%			&+ F^l_B)+ \gamma(K_i(F^l_i-\xi^l) + F^l_B)\cdot K_i(-K_p \text{grad}^l\phi-K_d\xi^l) \\
%		 =\;&(\alpha - \beta K_p) \text{grad}^l\phi\cdot\xi^l - \beta K_d \Vert \xi^l \Vert^2\\ 
%		 	&+ (\beta + \gamma K_i^2)\xi^l\cdot(K_iF^l_i + F^l_B)
%			- \gamma K_i \Vert K_iF^l_i + F^l_B\Vert^2\\
%		 =\;&(\alpha - \beta K_p) \text{grad}^l\phi\cdot\xi^l\\
%		 	&-\left(-\frac{1}{4\gamma K_i}\beta^2 + \left(K_d - \frac{1}{2}K_i\right)\beta-\frac{1}{4}\gamma K_i^3\right)\Vert\xi^l\Vert^2\\
%		 	&-\left\Vert\frac{1}{2}\frac{\beta+\gamma K_i^2}{\sqrt{\gamma K_i}}\xi^l - \sqrt{\gamma K_i}(K_iF^l_i + F^l_B)\right\Vert^2\;.
%\end{split}
%\end{equation}
We have $\dot{V} \leq 0$ if $P := -\frac{1}{4\gamma k_i}\beta^2 + \left(k_d - \frac{1}{2}k_i\right)\beta-\frac{1}{4}\gamma k_i^3>0$. $P$ is a position function of $\beta$ if
\begin{equation}\label{eq:BetaIdentifier}
\frac{\left(k_d-k_i/2\right)-\sqrt{\Delta}}{1/(2\gamma k_i)} < \beta < \frac{\left(k_d-k_i/2\right)+\sqrt{\Delta}}{1/(2\gamma k_i)}
\end{equation}
with $\Delta = k_d^2 - k_i k_d$ and we recover the condition $k_i < k_d$ to have $\Delta > 0$. Thus for any positive $k_p, k_d$ and $k_i < k_d$, we can find a Lyapunov function of the form \eqref{eq:2-nd Lyapunov function} and obtain $\dot V\leq 0$. The set where $\dot V = 0$ reduces to $\xi^l = 0$, hence $k_i F_i^l + F_B^l = 0$. To keep these conditions invariant, \eqref{eq:2-nd integral control force}\eqref{eq:2-nd integral term} impose $\text{grad}^l\phi=0$. The LaSalle invariance principle hence ensures convergence to the announced equilibrium set.

The arguments for stability only of $\phi=0$, and for the basin of attraction by making $\phi$ dominate the other terms of $V$, are the same as for Proposition 1.% Also for the basin of attraction we use the same argument, i.e.~we build on the fact that making $\alpha$ very large w.r.t.~$\beta,\gamma$ ensures that the $\phi$-term dominates the Lyapunov function. For fixed $K_d$ and $K_i$, the requirement \eqref{eq:BetaIdentifier} makes $\beta$ just proportional to $\gamma$; thus taking $K_p$ sufficiently large allows to get $\alpha = \beta K_p$ arbitrarily large w.r.t.~$\beta$ and $\gamma$.
\hfill $\square$\\
 
\subsection{Direct extensions}\label{ssec:extensions}

The previous section can of course be readily transposed to the case where both the inputs and the constant bias are on right-invariant rather than on left-invariant velocities/torques. We next analyze a system controlled on left-invariant inputs but with bias constant under right-invariant transport. We conclude with a brief discussion of the extension to underactuated systems.

%% LEFT & RIGHT MIXED
Let us first write the equations for the right-invariant case, e.g.~for a first-order system under PI control:
\begin{eqnarray}
\label{eq:Rg} R_{g^{-1}}\frac{d}{dt} g & = & -k_p \text{grad}^r \phi + k_i\xi^r_i + \xi^r_B\\
\nonumber \frac{d}{dt}\xi^r_i & = & -k_p \text{grad}^r \phi \;.
\end{eqnarray}
With the adjoint action (see Section \ref{ssec:Liegroups}), \eqref{eq:Rg} rewrites:
\begin{equation}
\label{eq:Rghop} 
\begin{split}
R_{g^{-1}}\frac{d}{dt} g \;& =\;  Ad_{g(t)}(-k_p \text{grad}^{l*} \phi + k_i\xi^l_i) + \xi^r_B \\
\frac{d}{dt} \xi^l_i \;& =\;   -k_p \text{grad}^{l*} \phi - [\xi^l, \xi^l_i] \; ,
\end{split}
\end{equation}
where $\text{grad}^{l*} \phi := Ad_{g^{-1}} Ad_{g^{-1}}^* \text{grad}^{l} \phi$. The second equation is obtained from $\frac{d}{dt}(Ad_{g(t)} \xi^l_i) =  Ad_{g(t)} (\tfrac{d}{dt}\xi^l_i + [\xi^l,\xi^l_i]) = Ad_{g(t)}(-k_p \text{grad}^{l*} \phi)$. Similarly, the second-order system with PID control
%\begin{equation}
%\label{eq:Rg2} 
%\begin{split}
%R_{g^{-1}}\frac{d}{dt}g \;& = \; \xi^r \\
%\frac{d}{dt}\xi^r \; & = \; -K_p \text{grad}^r \phi - K_d\xi^r + K_iF^r_i + F^r_B\\
%\frac{d}{dt}F^r_i \; & = \; -K_p \text{grad}^r \phi - K_d\xi^r \;.
%\end{split}
%\end{equation}
rewrites
\begin{equation}
\label{eq:Rg2hop}
\begin{split}
R_{g^{-1}}\frac{d}{dt}g \;& = \; \xi^r \\
\frac{d}{dt}\xi^r \;& = \; Ad_{g(t)}(-k_p \text{grad}^{l*} \phi - k_d\xi^l + k_iF^l_i) + F^r_B\\
\frac{d}{dt}F^l_i \; & = \; -k_p \text{grad}^{l*} \phi - k_d\xi^l - [\xi^l, F^l_i] \;.
\end{split} 
\end{equation}
After rewriting, the brackets behind $Ad_{g(t)}$ contain the control inputs, in left-invariant frame, and the respective last lines define the integral control computation, in left-invariant frame as well. The $g$-dependent change of frame induces a correction term in the latter. The last line of \eqref{eq:Rg2hop} can be implemented as such, but in the last line of \eqref{eq:Rghop} the factor $\xi^l = -k_p \text{grad}^{l*} \phi + k_i\xi^l_i + Ad_{g(t)}^{-1}(\xi^r_B)$ would contain the unknown bias. Therefore the actual controller must replace that equation by a best guess (note that $\bar{\xi}^l$ need not contain $k_i \xi^l_i$ since $[k_i \xi^l_i, \xi^l_i] = 0$ anyways):
\begin{equation}
\label{eq:xibar} 
\begin{split}
\frac{d}{dt} \xi^l_i \; & = \; -K_p \text{grad}^{l*} \phi - [\bar{\xi}^l, \xi^l_i] \; \text{ with}\\
\bar{\xi}^l \; & = \; -K_p \text{grad}^{l*} \phi \; .
\end{split}
\end{equation}

\noindent \textbf{Corollary 3:} The system \eqref{eq:Rg2hop} for a \emph{left-invariant-controlled} system with \emph{right-invariant constant bias} features the same convergence properties as the system in Proposition 2, with all left-invariant quantities replaced by the corresponding right-invariant quantities.\vspace{2mm}

\noindent \emph{Proof.} \eqref{eq:Rg2hop} is strictly equivalent to the verbatim transcription of \eqref{eq:2-nd integral control velocity}-\eqref{eq:2-nd integral term} from left-invariant to right-invariant.\hfill $\square$\\

\noindent \textbf{Proposition 4: (crossed PI control)} On a Lie group with \emph{unitary adjoint representation}, the system \eqref{eq:Rghop}\eqref{eq:xibar} for a \emph{left-invariant-controlled} system with \emph{right-invariant constant bias} features the same convergence properties as the system in Proposition 1.\vspace{2mm}

\noindent \emph{Proof.} The unitary adjoint representation is necessary in order to use the property $a \cdot [a,b] = a \cdot [b,a] = 0$ for some terms in the derivative of the Lyapunov function. It also implies $\text{grad}^{l*} \phi = \text{grad}^{l} \phi$ such that, using the identity $\tfrac{d}{dt}(Ad_g^{-1}) \omega = [\xi^l, Ad_g^{-1} \omega]$, the time derivative of $V = \alpha \phi + \tfrac{\beta}{2} \Vert k_i \xi_i^l + Ad_g^{-1} \xi^r_B \Vert^2$ finally reduces to $\dot{V} = - k_p \alpha \Vert Ad_{g^{-1}}^* \text{grad}^{l} \phi \Vert^2$ when $\alpha = k_p k_i \beta$. The rest of the proof is as for Proposition 1.\hfill $\square$\\

The additional condition for PI control comes from the replacement of $\xi^l$ by $\bar{\xi}^l$ in \eqref{eq:xibar}. With a unitary adjoint representation, the norms of a left-and right-invariant quantities are equal, therefore the same Lyapunov function can be taken for Prop.1 and Prop.4; we also get $\text{grad}^{l*} \phi = \text{grad}^{l} \phi$, in the control expression. However, while the group $SO(n)$ of rigid body rotations has a unitary adjoint representation, the group $SE(n)$ of rotations and translations does not (at least not for all velocities).\\

%% UNDERACTUATED
We now briefly discuss underactuated, more precisely nonholonomic systems. A typical nonholonomic constraint (e.g.~steering control~\cite{justh2004equilibria, justh2005natural}) restricts velocity to the affine space $\xi^l = a_0 + \sum_{j=1}^m \, a_j u_j$ for some fixed orthogonal $a_j \in \mathfrak{g}$, $j=0,1,...,m$ and input commands $u_j \in \mathbb{R}$, $j=1,2,...,m$. For $a_0 \neq 0$, the system features no steady state. Moreover the system is often not \emph{locally} controllable in practice and specific motion planning methods must be used for stabilization, unless the target is relaxed to a set. For $a_0 = 0$, a gradient-based proportional controller would in general be insufficient and feature undesirable invariant sets of dimension equal to the nonholonomic constraints.% whereas discussing second-order underactuated systems on Lie groups seems to make general sense only if we include inherent mechanical dynamics (e.g.~Euler equations for rotation).

Notwithstanding these issues, (set) stabilization can be obtained as a direct extension of proportional control in certain cases. But even then, adding integral control and biases may cause difficulties. For a left-invariant-controlled system with left-invariant-constant bias, as equations \eqref{eqmy1},\eqref{eqmy2} can be adapted, it is clear that the integral control can only cancel the bias if the latter also belongs to the actuated subspace. For a left-invariant-controlled system with right-invariant-constant bias, the integration of something like \eqref{eq:xibar} with $\text{grad}^{l*} \phi$ replaced by a velocity belonging to the actuated subspace would not guarantee that also $\xi_i^l$ belongs to the actuated subspace, due to the last term which reflects the change of frame. An implementable integral control should hence further project down the integral term, with further restrictions on the bias that it can cancel. Eventually, it is doubtful whether all these restr
 ictions 
 on the simple approach would cover a situation that is practically meaningful.

%%%%%%%%%%%% APPLICATIONS %%%%%%%%%%%%

\section{APPLICATION EXAMPLES}\label{applications} %AS: done1
We now illustrate our method on two robotic applications. Firstly we consider satellite attitude control on the rotation group $SO(3)$ and then we turn to complete 3-dimensional motion control of e.g.~an underwater vehicle on the group $SE(3)$ of rotations and translations. We assume that both systems are fully actuated, that is, the satellite can command rotations around any axis in 3-dimensional space, and the underwater vehicle can, in any situation, command translations along all 3 degrees of freedom and rotations around any axis in 3-dimensional space. We present simulation results for each case.

\subsection{Attitude control of a satellite}
Let $Q_s(t)\in SO(3), t\in\mathbb{R}_+$ denote the actual trajectory of a satellite's attitude and $Q_r(t)\in SO(3), t\in\mathbb{R}_+$ the target trajectory. A configuration error function can be defined by $\phi(Q)=\frac{1}{2} \text{tr}(I_{3\times 3}-Q)$ with
$Q = Q_r^T \, Q_s$, such that $\phi(Q)=0$ corresponds to our target $Q_s=Q_r$. The gradient of the configuration error function is
\begin{equation}
\text{grad}^l\phi \;=\; [\text{skew}(Q)]^\vee = [\tfrac{1}{2}(Q-Q^T)]^\vee
\end{equation}
in terms of angular velocity; the critical points of $\phi$ amount to $\phi(0)=0$ and the set of its maxima, where the satellite is turned by 180 degrees around some axis with respect to the target. For simplicity, we restrict the following to the case where $Q_r$ is constant, e.g.~$Q_r = I_{3\times 3}$ without loss of generality. The case of time-varying $Q_r(t)$ requires feedforward; explicitly, the effect of $\tfrac{d}{dt}Q_r = Q_r [\omega^l_{\chi}]^\wedge$ on the evolution of $Q = Q_r^T Q_s$ can be perfectly canceled by adding a feedforward angular velocity command $\omega_{ff}^l = Q^T \omega^l_{\chi}$ (see general expression in Section \ref{ssec:refi}) to the dynamics of $Q_s$.
%Let $\omega^l=[Q^T \dot Q]^\vee$ denote the body-fixed angular velocity of the satellite, which is also the velocity error.
Particularizing \eqref{eq:1-st integral control} and \eqref{eq:1-st integral term} gives the first order integral controller on $SO(3)$:
\begin{equation*}
\begin{split}
Q^{T}\frac{d}{dt} Q \;& = \; \left[ -k_p\text{skew}(Q)^\vee + k_i\omega^l_i + \omega^l_B \right]^\wedge
\,,\\
\frac{d}{dt}\omega^l_i \;& = \; -k_p\, [\text{skew}(Q)]^\vee \;.
\end{split}
\end{equation*}
where $\omega^l_i$ is the integral term and $\omega^l_B$ is a body-fixed bias.
Such bias might be caused by a mis-calibration of internal flywheels reference velocity.
According to Proposition 1, if we take $k_p$ or $k_i$ sufficiently large and avoid starting at orientations exactly opposite to the desired one, then this controller will stabilize $Q$ to the identity with $\omega^l = 0$, while $\omega_i^l$ gets the value $-\omega^l_B/k_i$.

\begin{figure}[thpb]
 	\centering
 	\makebox{\parbox{3.2in}{\includegraphics[scale=0.54]{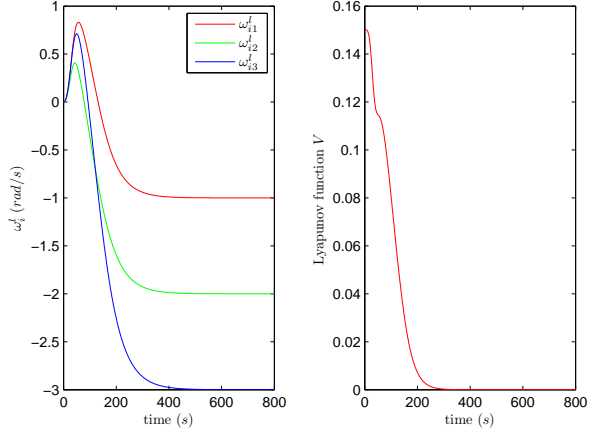}}}
 	\caption{Integral term and Lyapunov function for first-order system on $SO(3)$}
 	\label{SO(3)_1st_IntegralLyapunov}
\end{figure}

We have simulated this controller with arbitrary parameters $k_p=0.04$, $k_i=0.01$ and drift $\omega^l_B=0.01\,[1,2,3]^T$, starting from $Q(0)=\frac{1}{3}\,[-1~2~2;2~-1~2;2~2~-1]$ and $\omega^l_i(0)=[0,0,0]^T$. The evolution of the integral term and of the decreasing Lyapunov function (with $\alpha=0.04$ and $\beta=100$) are shown on Fig.\ref{SO(3)_1st_IntegralLyapunov}.

% % %  simulation results of first-order system on SO(3) % % %

\begin{figure}[thpb]
 	\centering
 	\makebox{\parbox{3.2in}{\includegraphics[scale=0.54]{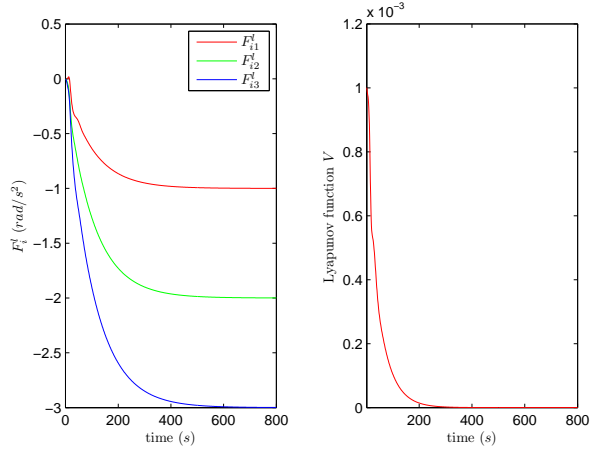}}}
 	\caption{Integral term and Lyapunov function for second-order system on $SO(3)$}
 	\label{SO(3)_2nd_IntegralLyapunov}
\end{figure}

Similarly, the second-order dynamical controller \eqref{eq:2-nd integral control velocity}-\eqref{eq:2-nd integral term} yields for satellite attitude the PID-controlled system:
\begin{equation*}
\begin{split}
Q^T\frac{d}{dt}Q \;& = \; [\omega^l]^\wedge \\
\frac{d}{dt}\omega^l \;& = \; -k_p [\text{skew}(Q)]^\vee - k_d\omega^l + k_iF^l_i + F^l_B\\
\frac{d}{dt}F^l_i \;& = \; -k_p [\text{skew}(Q)]^\vee - k_d\omega^l \;.
\end{split}
\end{equation*}
From Proposition 2, the configuration $Q$ under this torque control will converge to the identity and stay there, while the torque bias is asymptotically countered by $F^l_i=-F^l_B/k_i$. This is illustrated in the simulation reported on Fig.~\ref{SO(3)_2nd_IntegralLyapunov}, which was made with the same initial values (at rest) as for the first-order case, the same bias but now on the torque $F^l_B=0.01\,[1,2,3]^T$, and parameters $k_p=0.04$, $k_i=0.01$, $k_d=0.2$.
A torque bias could in practice result from leakage in jet-actuated control.
The choice $\alpha=0.04*0.0039$, $\beta=0.0039$, $\gamma=1$ satisfies \eqref{eq:BetaIdentifier} and other sufficient conditions for a decreasing Lyapunov function.

% % %  simulation results of second-order system on SO(3) % % %

\subsection{Full control of an underwater vehicle}
We next consider a vehicle with not only rotations $Q \in SO(3)$ but also translations $p \in \mathbb{R}^3$, to form a configuration $g\in SE(3)$. Again we assume a setup where the target is the group identity $p=0$ and $Q=I_{3\times 3}$. The error function on $SE(3)$ is:
\begin{equation}
\phi_1(Q,p)=\frac{1}{2} \text{tr}(I_{3\times 3}-Q)+\frac{1}{2}\parallel p\parallel^2\;.
\end{equation}
Then most computations follow directly from the ones for $SO(3)$. E.g.~the critical points of $\phi_1$ are at $p=0$ with either $Q =I_{3\times 3}$ or $Q$ describing any 180 degree rotation; the latter is a saddle point set where $\phi_1 = 2$. Introducing a small weighting factor in front of $\parallel p\parallel^2$ would allow to arbitrarily increase the domain of translations $p$ that are included in the basin of attraction where $\phi_1 < \phi_{1c}$.

\begin{figure}[thpb]
 	\centering
 	\makebox{\parbox{3.2in}{\includegraphics[scale=0.54]{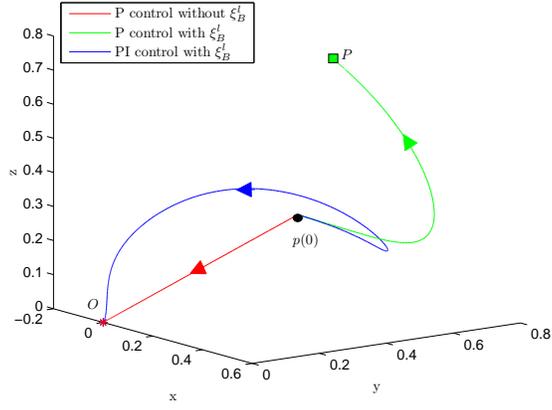}}}
 	\caption{Trajectories of different control strategies for first-order system on $SE(3)$ --- only the position part of $g$ is represented}
 	\label{SE(3)_1st_Trajectory}
\end{figure}

\begin{figure}[thpb]
 	\centering
 	\makebox{\parbox{3.2in}{\includegraphics[scale=0.54]{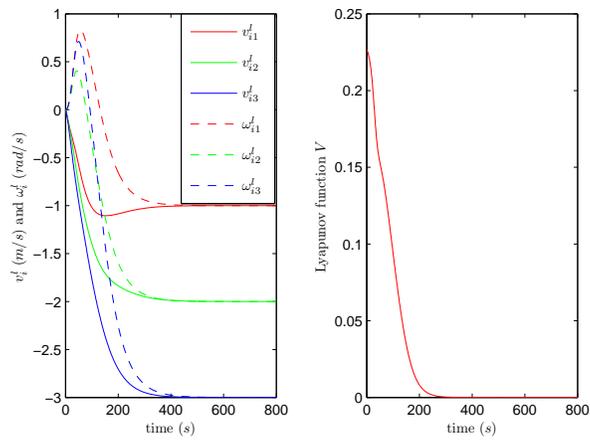}}}
 	\caption{Integral term and Lyapunov function for first-order system on $SE(3)$}
 	\label{SE(3)_1st_IntegralLyapunov}
\end{figure}

The velocity in $\mathfrak{se}(3)$ comprises 3-dimensional rotation velocity $\omega^l$ and 3-dimensional translation velocity $v^l$, both in body frame. Translation and rotation are coupled as explained in Section \ref{ssec:Liegroups} and with that matrix notation the PI-controlled system \eqref{eq:1-st integral control},\eqref{eq:1-st integral term} rewrites:
\begin{eqnarray*}
g^{-1}\frac{d}{dt} g & = & -k_p\begin{bmatrix}
\text{skew}(Q) & Q^T p \\ 0 & 0
\end{bmatrix} \\
& & + k_i\begin{bmatrix}
[\omega^l_i]^\wedge & v^l_i \\ 0 & 0
\end{bmatrix} + \begin{bmatrix}
[\omega^l_B]^\wedge & v^l_B \\ 0 & 0
\end{bmatrix}\,,\\
\frac{d}{dt}\begin{bmatrix}
\omega^l_i\\
v^l_i
\end{bmatrix} & = & -k_p\begin{bmatrix}
[\text{skew}(Q)]^\vee\\
Q^T p
\end{bmatrix} \;.
\end{eqnarray*}
%where $[\omega^l_i,v^l_i]^T$ and $[\omega^l_B,v^l_B]^T$ are respectively the integral term and the bias in body frame. 
Simulation results are shown in Fig.~\ref*{SE(3)_1st_Trajectory}, representing only the position part of $g$. In addition to the parameters already used for rotation, we take $v^l_B=0.01\,[1,2,3]^T$, $p(0)=\frac{1}{3}\,[1,1,1]^T$ and $v_i(0)=[0,0,0]^T$. We have plotted the ideal trajectory of P control \emph{without bias} as a reference. In presence of bias, under \emph{P control} the position moves in a wrong direction, converges to a stable point $P=[0.25,0.5,0.75]^T$ and stays there with a steady-state error whose gradient pull compensates the bias. Under \emph{PI control}, the bias still starts the system in the wrong direction, but once the integral term takes over it converges back to the desired equilibrium $O=[0,0,0]^T$\, and stays there, while the bias is countered by an integral term $k_i \xi^l_i=-\xi^l_B$ (see Fig.~\ref*{SE(3)_1st_IntegralLyapunov}). 

% % %  simulation results of first-order system on SE(3) % % %

The second-order case follows exactly the same principles. Simulations can be easily established with the corresponding parameters taken over from previous cases. In accordance with Proposition 2, the system shall converge to the equilibrium where $Q = I_{3\times 3}$ and $p=0$, while the bias in actuators is countered by $F^l_{\omega i} =-\frac{1}{k_i}F^l_{\omega B}$ and $F^l_{vi}=-\frac{1}{k_i}F^l_{vB}$\;.

Besides calibration errors or actuator leakage, a bias on the underwater vehicle could be caused by slow (errors in cancellation of) internal dynamics. Also a constant bias in \emph{inertial} frame would make sense, e.g.~caused by ocean flow (see extensions Section \ref{ssec:extensions}). The second-order system is then covered by Corollary 3, but for the first-order model $SE(3)$ does not satisfy the requirement of unitary adjoint representation for Proposition 4. Realistic settings also include the nonholonomic ``steering control'' case, which is worth future interest.

%%%%%%%%%%%% CONCLUSIONS %%%%%%%%%%%%

\section{CONCLUSIONS}\label{conclusion} %AS: done2
We propose a general integral control method for systems on nonlinear manifolds by explicitly defining the integral term as the integration of the control commands in the corresponding transported tangent spaces. In particular, for Lie groups, the transport maps associated to left and right group actions are a natural choice. Under this rigorous definition, we can easily extend PID control from Euclidean space to Lie groups. Both first order integrators with bias in velocity and second order integrators with bias in controlled torque are shown to be well corrected by applying our integral control. Stability is proved with Lyapunov functions. We also take typical applications in robotics as examples to illustrate the physical meanings of the setting and developments, and to confirm the stability in simulation results. As for linear systems, the potential advantage of PID control over the observer-based approach to bias rejection on Lie groups is that PID controllers do not have to simulate and hence know the full dynamical model of the system. For instance, it can be expected that the stability proven here on simple examples remains valid more or less verbatim if actuator dynamics are added to the system. Future research should investigate to which underactuated contexts the approach could be adapted, especially when left-invariant control (i.e.~inputs constrained in body frame) is combined with right-invariant constant biases (i.e.~forces/torques/flows attached to inertial frame). Another opened research direction is more explicit integral control for Riemannian manifolds, i.e.~investigating meaningful transport maps both for applications and regarding convergence properties. In this regard, we already note that the equivalent of a ``constant bias'' cannot be defined on all manifolds, as e.g.~the even-dimensional spheres cannot support non-vanishing smooth vector fields~\cite[Th.2.2.2]{burns2005differential}. The implications of our integral controller for robust coordinated motion should also be investigated.

\section*{Acknowledgment}

This paper presents research results of the Belgian Network DYSCO  
(Dynamical Systems, Control, and Optimization), funded by the  
Interuniversity Attraction Poles Programme, initiated by the Belgian  
State,  Science Policy Office. The first author's visit to the SYSTeMS research group is supported by CSC-grant (No.201306740021) and the associated BOF-cofunding, initiated by the China Scholarship Council and Ghent University respectively.

\bibliography{icol}
%\end{multicols}

\end{document}